\documentclass[a4paper,fleqn,usenatbib]{mnras}

\usepackage{mathptmx}

\usepackage[T1]{fontenc}
\usepackage{ae,aecompl}

\usepackage{graphicx}	
\usepackage{amsmath}	
\usepackage{amssymb}	
\usepackage{units}
\usepackage{tikz}
\usetikzlibrary{decorations.pathmorphing}
\usetikzlibrary{decorations.markings}

\newcommand{\lsi}{LS~I~$+61^{\circ}303$}
\newcommand{\fermilat}{\textit{Fermi}-LAT}
\newcommand{\gammaray}{$\gamma$-ray}
\newcommand{\Xing}{X17}

\title[
  OVRO and \fermilat{} monitoring of \lsi{}
]{
  Simultaneous long-term monitoring of \lsi{} by OVRO and \fermilat{}
}

\author[
  F. Jaron et al.
]{
  Fr\'ed\'eric Jaron,$^{1}$\thanks{E-mail: fjaron@mpifr-bonn.mpg.de}
  Maria Massi,$^{2}$
  Sebastian Kiehlmann,$^{3}$
  Talvikki Hovatta$^{4}$
  \\
  $^{1}$Institute of Geodesy and Geoinformation -- IGG, University of Bonn,
  Nu{\ss}allee 17,
  D-53115 Bonn,
  Germany\\
  $^{2}$Max Planck Institute for Radio Astronomy,
  Auf dem H\"ugel 69,
  D-53121 Bonn,
  Germany\\
  $^{3}$Owens Valley Radio Observatory,
  California Institute of Technology, Pasadena,
  CA 91125,
  USA\\
  $^{4}$Tuorla Observatory,
  University of Turku,
  V\"ais\"al\"antie 20,
  FI-21500 Piikki\"o,
  Finland
}

\date{Accepted XXX. Received YYY; in original form ZZZ}

\pubyear{2018}

\begin{document}
\label{firstpage}
\pagerange{\pageref{firstpage}--\pageref{lastpage}}
\maketitle

\begin{abstract}
  Previous long-term monitorings of the \gammaray{}-loud X-ray binary \lsi{} have revealed the presence of a long-term modulation of $\sim 4.5$\,years. After nine years of simultaneous monitoring of \lsi{} by the Owens Valley Radio Observatory and the \fermilat{}, two cycles of the long-term period are now available. Here we perform timing-analysis on the radio and the \gammaray{} light curves. We confirm the presence of previously detected periodicities at both radio and GeV \gammaray{} wavelengths. Moreover, we discover an offset of the long-term modulation between radio and \gammaray{} data which could imply different locations of the radio (15 GHz) and GeV emission along the precessing jet.
\end{abstract}




\section{Introduction} \label{sect:introduction}

The stellar system \lsi{} is detected all over the electromagnetic spectrum ranging from radio to very high energy \gammaray{}s \citep{Taylor1982, Paredes1994, Mendelson1989, Zamanov1999, Harrison2000, Abdo2009, Albert2006}.  The binary consists of a Be type star \citep{Casares2005} and a compact object in an eccentric orbit. Optical polarization observations have determined a value of 25~degrees for the position angle of the rotational axis of the Be disk in \lsi{} \citep{Nagae2006}. Assuming that the inclination angle of the orbit coincides with the position angle of the rotational axis of the Be star, for an inclination angle of 25 degrees the compact object would be a black hole \citep[e.g.,][]{Casares2005}. Indeed the X-ray characteristics of \lsi{} fit well those of accreting black holes \citep{Massi2017}, and not those of a radio pulsar, which is the alternative suggested model for \lsi{} \citep[e.g.,][]{Dubus2006}.

By Bayesian analysis of 20~years of radio data \citet{Gregory2002} determined the orbital period, $P_1 = 26.4960 \pm 0.0028$\,d. Lomb-Scargle timing analysis \citep{Lomb1976, Scargle1982} of 37~years of radio data resulted in the detection of a second period $P_2 = 26.935 \pm 0.013$\,d \citep{Massi2016} consistent with a previously determined precession period of the radio jet mapped in VLBI images \citep{Massi2012}. Recently VLBA astrometry increased the accuracy to $P_2=26.926 \pm 0.005$\,d \citep{Wu2017}.

The presence of a very stable long-term flux modulation affecting the radio emission \citep{Massi2016} finds a straightforward explanation as being the result of the beating between the two intrinsic and stable orbital and precession periodicities $P_1$ and $P_2$ \citep{Massi2013}. For $P_1 = 26.496$\,d and the last determination by \citet{Wu2017} of $P_2 = 26.926$\,d the beating (i.e., $(\nu_1 - \nu_2)^{-1}$) results in $P_{\rm long} = 1659$\,d. This hypothesis was tested by \citet{Massi2014} who developed a physical model (based on the work of \citealt{Kaiser2006}) of a self-absorbed jet, periodically ($P_1$) refilled with electrons, and which precesses with a period of $P_2$ and thus continuously changes angle with respect to the line of sight of an observer, giving rise to periodic changes in the Doppler boosting of the intrinsic emission. The model by \citet{Massi2014} reproduces the observed radio light curve over the last four decades. The alternative scenario in which the long-term modulation could be the result of changes in the Be star wind is unlikely. The unstable quasi-periodic behavior of Be stars has been reviewed by \citet{Rivinius2013}, and we refer to the introduction of \citet{Jaron2017} for a discussion of this issue in comparison to the very stable periodic behavior of \lsi{}.

In the GeV regime (as continously monitored by the Large Area Telescope onboard the \textit{Fermi} satellite, i.e., \fermilat{}, \citealt{Atwood2009}), the orbital period was clearly detected first by \citet{Abdo2009}. However, as found out by \citet{Ackermann2013}, the behavior around periastron differs from that around apastron in the way that the long-term modulation only affects the latter and not the former. In agreement with these findings, \citet{Jaron2014} report the GeV emission around apastron to be further modulated by the precession period $P_2$, known from the radio emission. By extending the physical model of \citet{Massi2014} to the GeV regime, including both synchrotron self-Compton and external inverse Compton, \citet{Jaron2016} reproduce the radio and GeV light curves and their different timing characteristics at periastron and apastron.

While the stability of the two periodicities $P_1$ and $P_2$ could be well demonstrated for the radio regime on the basis of almost 40 years of data, corresponding to more than eight cycles of the long-term modulation, continuous monitoring of \lsi{} in the GeV regime has only been ongoing since 2008. The now completed two long-term cycles of \fermilat{} observations and the availability of almost simultaneous radio monitoring by the Owens Valley Radio Observatory (OVRO) allows us to investigate on different influences of the two periodicities $P_1$ and $P_2$, and their beating, in the radio and GeV emission.

\section{Observations}

\begin{figure*}
  \includegraphics{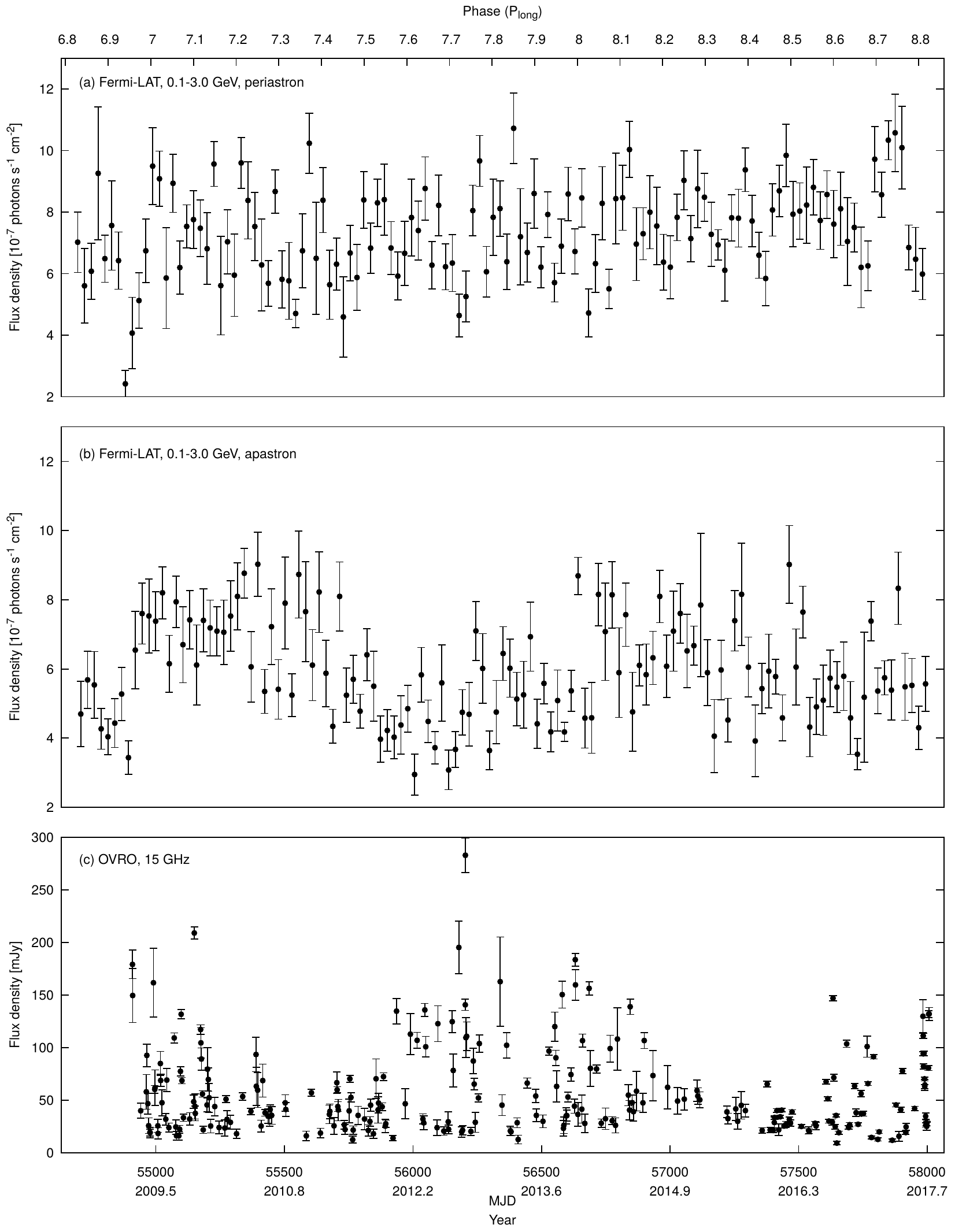}
  \caption{
    Light curves used for the analysis presented here, aligned in time. Observation time is expressed as MJD and year in the lower $x$-axis of the bottom panel and in terms of long-term phase in the upper $x$-axis of the upper panel.
    (a) \fermilat{}, energy range 0.1--3\,GeV, averaged over one orbit, periastron ($\Phi = 0.0-0.5$) data,
    (b) apastron data ($\Phi = 0.5-1.0$).
    (c) Radio data at 15\,GHz obtained by OVRO monitoring. Only data above $3\,\sigma$ have been selected for the analysis and are plotted here.
    Between MJD~56400 and 56550 the apparent dip is due to lack of data, in fact there are only seven data points during this interval (see Sect.~\ref{sec:obsOVRO}).
  }
  \label{fig:lightcurves}
\end{figure*}

\subsection{OVRO} \label{sec:obsOVRO}

The OVRO has been monitoring \lsi{} at 15\,GHz since 2009~March~18 (MJD~54908) with an average cadence of 5.8~days, i.e., well above the Nyquist limit of the periodicities of months to years considered here. The telescope uses off-axis dual-beam optics and a cryogenic receiver with 3~GHz bandwidth centred at 15~GHz. Atmospheric and ground contributions as well as gain fluctuations are removed with the double switching technique \citep{Readhead1989} where the observations are conducted in an ON-ON fashion so that one of the beams is always pointed on the source. Until May 2014 the two beams were rapidly alternated using a Dicke switch, since May 2014 when a new pseudo-correlation receiver replaced the old receiver a 180~degree phase switch is used.
Relative calibration is obtained with a temperature-stable noise diode to compensate for gain drifts. The primary flux density calibrator is 3C~286 with an assumed value of 3.44~Jy \citep{Baars1977}, DR21 is used as secondary calibrator source. Details of the observation and data reduction schemes are given in \citet{Richards2011}.

In order to increase the signal to noise ratio, we only include data above $3\sigma$ in the analysis. In the bottom panel of Fig.~\ref{fig:lightcurves} the light curve resulting from these observations is plotted. Covering a long-term phase interval of $\Theta = 6.9 - 8.9$ (for $P_{\rm long} = 1659$\,d, see second $x$-axis in the upper panel) two cycles of the long-term modulation have now been monitored by OVRO. The long-term modulation is clearly visible in this plot. During the maximum of the long-term modulation there is a time interval (MJD 56400--56550) of 140\,d (i.e., six orbital cycles of $\sim 26.5$\,d) very poorly sampled: There are only seven observations, which in addition do not coincide with radio outbursts as predicted by \citet{Jaron2013}.

\subsection{\fermilat{}}

The \fermilat{} data used for the analysis presented here cover the time from MJD 54682--58015 (i.e., August 4 2008 until September 19 2017). In order to compute a light curve from the Pass~8 photon data downloaded from the \fermilat{} data server\footnote{\url{https://fermi.gsfc.nasa.gov/cgi-bin/ssc/LAT/LATDataQuery.cgi}} we include all data within 10$^{\circ}$ around the position of \lsi{} and use version v10r0p5 of the Fermi ScienceTools\footnote{Available from \url{https://fermi.gsfc.nasa.gov/ssc/data/analysis/software/}} to fit the source with a log-parabola,
\begin{equation}
  \frac{{\rm d}N}{{\rm d}E} = N_0\left(\frac{E}{E_{\rm b}}\right)^{-(\alpha + \beta\log(E/E_{\rm b}))},
\end{equation}
with all parameters left free for the fit. All parameters of all other sources within 10$^{\circ}$ are left free as well. Sources between 10 and 15$^{\circ}$ are included in the analysis with their parameters fixed to the catalog values. The diffuse emission was modeled using the Galactic model file gll\_iem\_v06.fits, and the isotropic spectral template iso\_P8R2\_SOURCE\_V6\_v06.txt. We perform this fit, restricted to the energy range $E = 0.1 - 3$\,GeV, for every time bin of width one day.

\section{Results}

\subsection{Timing analysis}

We perform Lomb-Scargle timing analysis \citep{Lomb1976, Scargle1982} on the OVRO and \fermilat{} data, using the UK Starlink software package in the same way as described in \citet{Massi2013}.

The orbital phase $\Phi$ of \lsi{} is defined as
\begin{equation}
  \Phi = \frac{t - t_0}{P_1} - {\rm int}\left(\frac{t - t_0}{P_1}\right), 
\end{equation}
where $t_0 = 43366.275$\,MJD \citep{Gregory2002}. Periastron occurs in this case at $\Phi = 0.23$ \citep{Casares2005}, i.e., not at zero orbital phase as is usual for binary systems. The long-term phase $\Theta$ is defined as
\begin{equation}
  \Theta = \frac{t - t_0}{P_{\rm long}} - {\rm int}\left(\frac{t - t_0}{P_{\rm long}}\right).
\end{equation}

\begin{table}
  \centering
  \begin{tabular}{l}
    \hline
    \hline
    \fermilat{}\\
    Periastron\\
    $P_1$ = $\unit[26.43 \pm 0.05]{d}$\\
    Apastron\\
    $P_1$ = $\unit[26.45 \pm 0.05]{d}$\\
    $P_2$ = $\unit[26.99 \pm 0.05]{d}$\\
    $P_{\rm long}$ = $\unit[1659 \pm 211]{d}$\\
    \hline
    OVRO\\
    $P_1$ = $\unit[26.49 \pm 0.06]{d}$\\
    $P_2$ = $\unit[26.95 \pm 0.06]{d}$\\
    $P_{\rm long}$ = $\unit[1771 \pm 258]{d}$\\
    \hline
    \hline
  \end{tabular}
  \caption{
    Periods found by the Lomb-Scargle timing analysis.
  }
  \label{tab:results}
\end{table}

\subsubsection{OVRO}

\begin{figure}
  \includegraphics{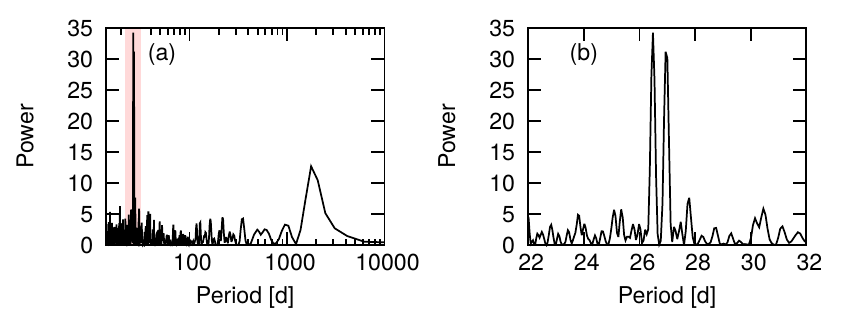}
  \caption{
    Lomb-Scargle timing analysis of the OVRO data.
    (a) The strongest feature is at $P_1$. $P_{\rm long}$ is present but with only one third of the power of $P_1$.
    (b) Entire data, zoom into the red-shaded area of panel a. Two peaks are well detected, i.e., the orbital period $P_1$ and the precession period~$P_2$.
  }
  \label{fig:OVRO_SC}
\end{figure}

The results of the timing analysis of the OVRO data are shown in Fig.~\ref{fig:OVRO_SC}. In the entire observed light curve (panel a) there is a clear signal at the orbital period and the long-term period. The zoom on the orbital period in panel b shows two peaks, i.e., the orbital period $P_1$ and the precession period $P_2$. The values of the detected periods are listed in the lower part of Table~\ref{tab:results}.

\subsubsection{\fermilat{}}

\begin{figure*}
  \includegraphics{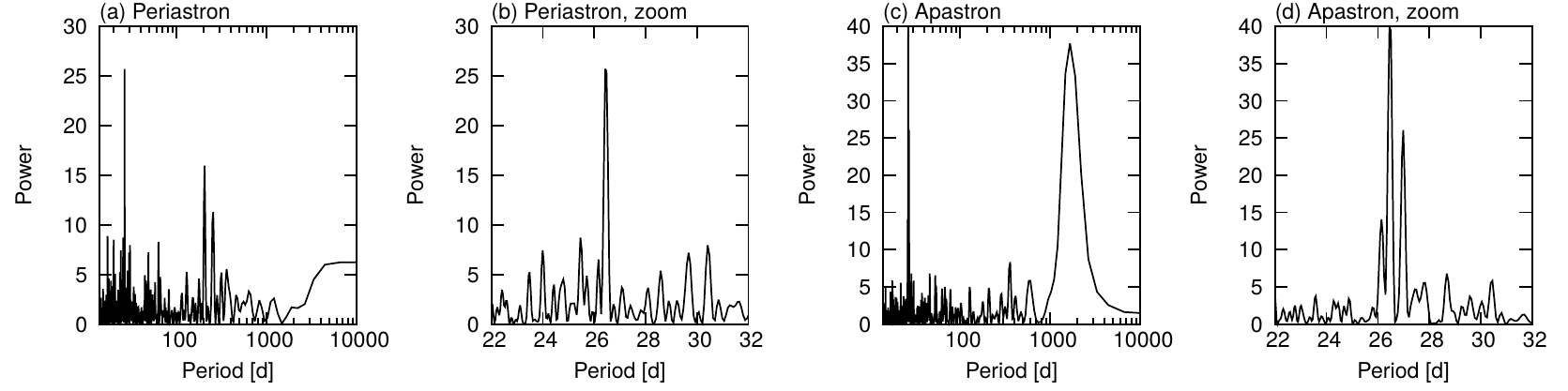}
  \caption{
    Timing analysis of the \fermilat{} data. Presented are Lomb-Scargle periodograms for different subselections of the dataset. Periastron is defined as orbital phase interval $\Phi = 0.0 - 0.5$, and apastron as $\Phi = 0.5 - 1.0$. 
    (a) Periastron. There is a peak at $P_1$.
    (b) Periastron, zoom. The only feature is a peak at $P_1$.
    (c) Apastron. A strong peak at the position of the long-term modulation appears.
    (d) Apastron, zoom. Beside the peak at $P_1$ there is a second peak at $P_2$. Both result highly significant in the randomisation tests (see \citealt{Massi2013} for details).
  }
  \label{fig:FermiLAT_SC}
\end{figure*}

Since the GeV emission from \lsi{} has different timing characteristics at periastron and apastron \citep{Ackermann2013, Jaron2014, Jaron2016}, we divided the light curve into $\Phi = 0.0-0.5$ (periastron), and $\Phi = 0.5-1.0$ (apastron).

The Lomb-Scargle periodograms for the \fermilat{} data are presented in Fig.~\ref{fig:FermiLAT_SC}. Panel~a and the zoom in b show a peak at the orbital period $P_1$. Panel~c, showing the result for the apastron data, contains a well pronounced peak at the position of the long-term period, and the zoom in panel~d shows two significant peaks, at the orbital period~$P_1$ and precession period~$P_2$. The values of the detected periods can be found in the upper part of Table~\ref{tab:results}.

\subsection{Folding the data}

All of the periods found by the timing analysis, presented in Table~\ref{tab:results}, are in agreement with the values reported before in the literature (see Sect.~\ref{sect:introduction}). For the folding of the radio and \gammaray{} light curves we use the following values: $P_1 = 26.4960$\,d \citep{Gregory2002} for the orbit, $P_2 = 26.926$\,d \citep{Wu2017} for the precession, and  
\begin{equation}
  P_{\rm long} = \frac{1}{\frac{1}{P_1} - \frac{1}{P_2}} = \unit[1659]{d},
\end{equation}
resulting as the beat period of the orbital and precession periods.

\subsubsection{OVRO}

\begin{figure}
  \includegraphics{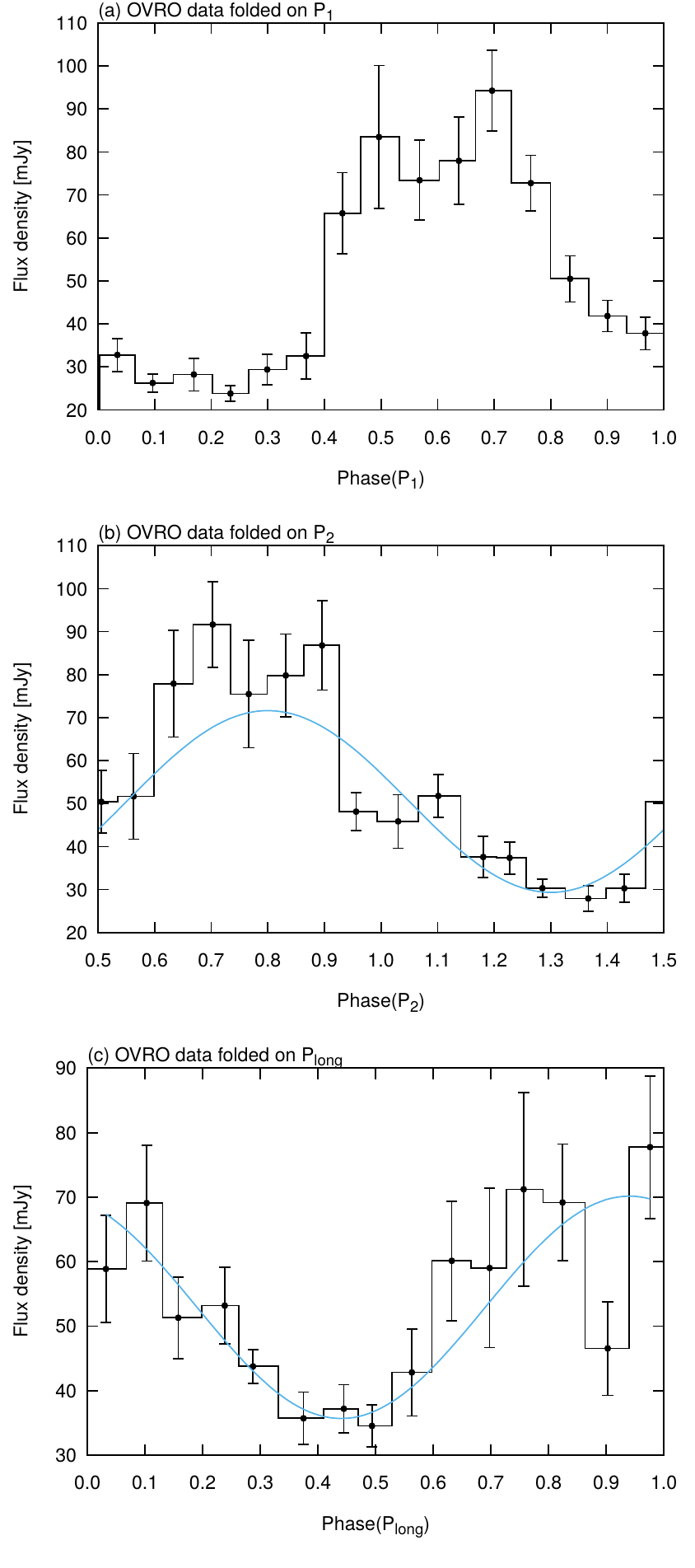}
  \caption{
    OVRO data folded.
    (a) Folded on $P_1$. The data cluster around orbital phase $\Phi = 0.6$, in agreement with the results from the long-term GBI monitoring (cf.\ the upper left panel of Fig.~4 in \citealt{Massi2013}).
    (b) Folded on $P_2$. The data cluster around ``precessional phase'' 0.8, again in agreement with the results from the long-term GBI monitoring (cf.\ the upper right panel of Fig.~4 in \citealt{Massi2013}).
    (c) Folded on $P_{\rm long}$. The long-term modulation is well visible, having a minimum around $\Theta  = 0.45$. The dip at $\Theta = 0.9$ is an artefact of the sampling, as explained in Sect.~\ref{sec:obsOVRO}.
  }
  \label{fig:OVRO_fold}
\end{figure}

Folding the OVRO data on $P_1$ results in the plot shown in Fig.~\ref{fig:OVRO_fold}\,a, the radio emission peaking from orbital phases $\Phi = 0.4-0.9$, i.e., around apastron, while at periastron the radio flux reaches only $\sim 30$\,mJy. This is in agreement with earlier observations at different radio frequencies \citep[][and references therein]{Massi2016}. The radio data folded with $P_2$ and $P_{\rm long}$ are shown in Figs~\ref{fig:OVRO_fold}\,b and c, respectively. The peak for the radio data folded with $P_2$ is at $\sim 0.8$ whereas radio data folded with $P_{\rm long}$ have a minimum at 0.45, i.e., maximum at 0.95.

\subsubsection{\fermilat{}}

\begin{figure}
  \includegraphics{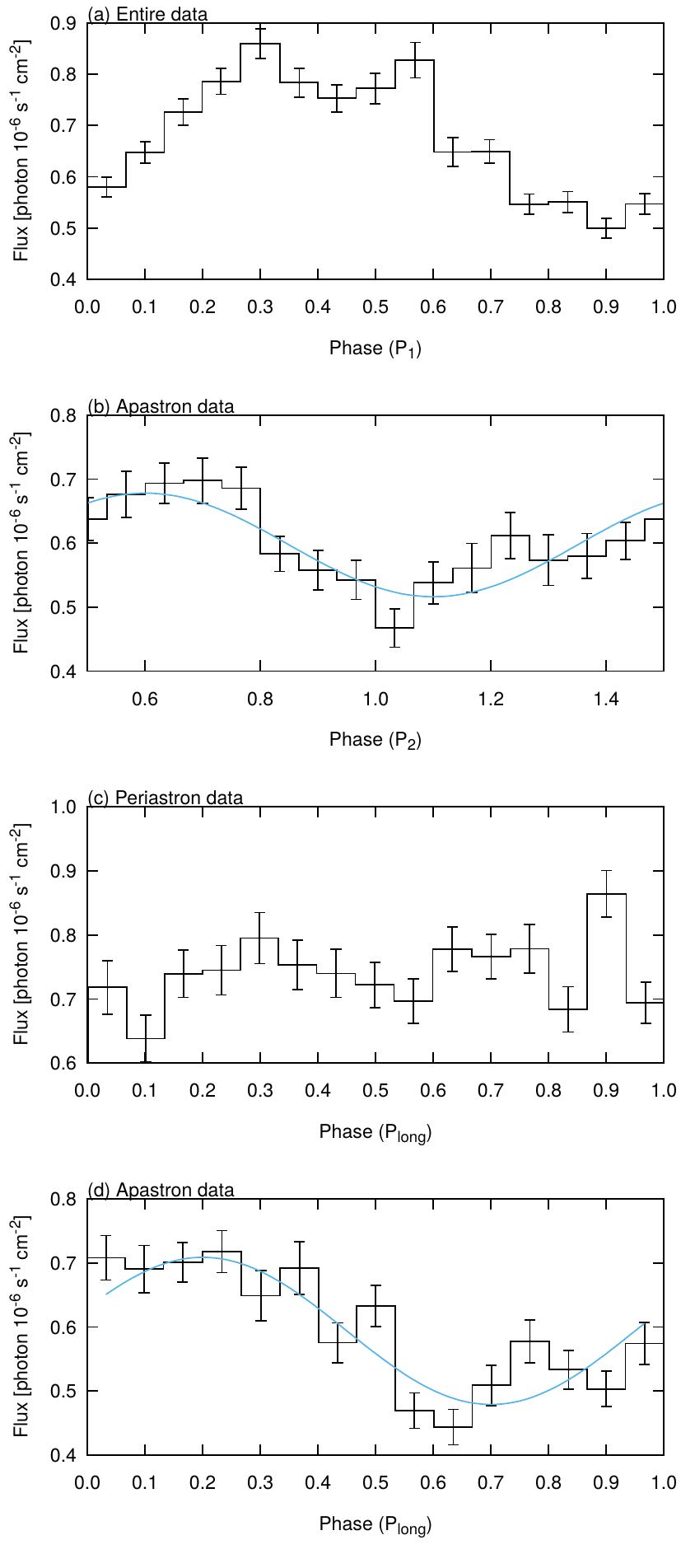}
  \caption{
    \fermilat{} data folded.
    (a) The entire dataset folded on the orbital period $P_1$. The modulation has the shape of a peak at periastron with an additional feature towards apastron.
    (b) Apastron ($\Phi = 0.5 - 1.0$) folded on $P_2$, revealing the shape of the flux modulation caused by the precession of the jet.
    (c) Periastron ($\Phi = 0.0 - 0.5$) folded on $P_{\rm long}$. No significant modulation.
    (d) Apastron ($\Phi = 0.5 - 1.0$) folded on $P_{\rm long}$. A significant modulation is well visible. 
  }
  \label{fig:FermiLAT_fold}
\end{figure}

\begin{figure}
  \includegraphics{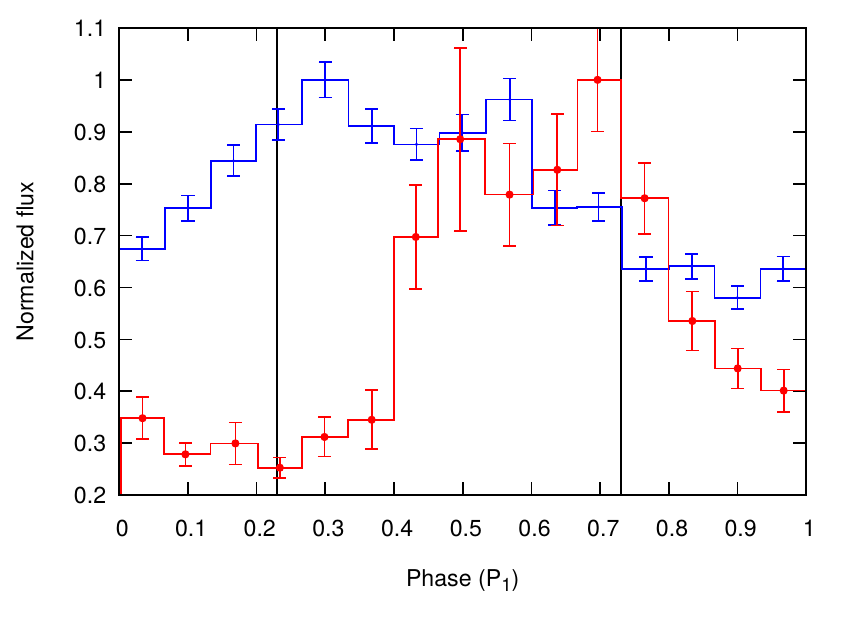}
  \caption{
    \fermilat{} (blue) and OVRO (red) data folded on $P_1$. Both datasets are normalized to their maximum values for a better comparability. Radio emission dominates at $\Phi = 0.4-0.8$. Gamma-ray emission shows in addition emission at periastron. The two vertical lines mark the orbital phases of periastron (0.23) and apastron, respectively.
  }
  \label{fig:FermiOVROfoldP1}
\end{figure}

In Fig.~\ref{fig:FermiLAT_fold}\,a the GeV data are folded on the orbital period\ $P_1$, showing that the emission covers the whole orbit. Figure~\ref{fig:FermiOVROfoldP1} shows the overlay with the radio emission, both normalized to their maximum fluxes for a better comparability of the two. In this plot it is evident that the radio emission is confined to orbital periods around apastron while the GeV emission covers the entire orbit.

Figures~\ref{fig:FermiLAT_fold}\,b and d present the apastron data folded on the precession period $P_2$ and the long-term period $P_{\rm long}$, respectively. As already pointed out by \citet{Ackermann2013} the GeV emission is only affected by the long-term modulation around apastron, and indeed only the plot in panel d shows a significant modulation for $P_{\rm long}$, while the plot in Fig.~\ref{fig:FermiLAT_fold}\,c does not show any significant modulation. The peak of $P_{\rm long}$ in the folded apastron data is at $\Theta \sim 0.2$ whereas the peak for apastron data folded with $P_2$ is at $\Phi(P_2) \sim 0.6$.

\subsubsection{Phase-offsets between radio and GeV}

\begin{table*}
  \begin{tabular}{l|llll}
    \hline
    \hline
    & $A$ & $B$ & $\phi_0$ & $\chi^2$\\
    \hline
    $P_2 = 26.926$\,d & & & \\
    \fermilat{} &
    $(8.09 \pm 1.22)\,10^{-8}$ &
    $(5.97 \pm 0.08)\,10^{-7}$ &
    $0.35 \pm 0.02$ &
    1.09\\
    OVRO &
    $(21.14 \pm 2.06)\,10^{-3}$ &
    $(50.47 \pm 1.69)\,10^{-3}$ &
    $0.55 \pm 0.02$ &
    2.40\\
    \hline
    $P_{\rm long} = 1659$\,d & & & \\
    \fermilat{} &
    $(1.15 \pm 0.18)\,10^{-7}$ &
    $(5.94 \pm 0.08)\,10^{-7}$ &
    $0.95 \pm 0.02$ &
    2.12\\
    OVRO &
    $(14.59 \pm 2.32)\,10^{-3}$ &
    $(51.37 \pm 1.79)\,10^{-3}$ &
    $0.69 \pm 0.02$ &
    1.17\\    
    \hline
    \hline
  \end{tabular}
  \caption{
    Parameters resulting from fitting the folded data with sine functions with amplitude $A$, constant offset $B$, and phase-offset $\phi_0$.
  }
  \label{tab:fitsin}
\end{table*}

When folding the data on the periods $P_2$ and $P_{\rm long}$, as presented in Figs~\ref{fig:OVRO_fold}\,b and c for the OVRO data, and in Figs~\ref{fig:FermiLAT_fold}\,b and d for the \fermilat{} apastron data, there is an offset between the otherwise similar shapes of the two. Aimed at investigating this phase-offset we fit the folded data with a sine function of the form
\begin{equation}
  f(\phi) = A\sin2\pi\left(\phi - \phi_0\right) + B.
\end{equation}
The parameters resulting from this fit are presented in Table~\ref{tab:fitsin} and the corresponding curves are plotted as blue solid lines in Figs~\ref{fig:OVRO_fold}\,b and c, and Figs~\ref{fig:FermiLAT_fold}\,b and d. The phase-offset is then given by
\begin{eqnarray}
  \Delta\phi_0(P_2) & = & \phi_{0,\,\rm radio} - \phi_{0,\,\rm GeV} = 0.20 \pm 0.03,\\
  \Delta\phi_0(P_{\rm long}) & = & \phi_{0,\,\rm radio} - \phi_{0,\,\rm GeV} = -0.26 \pm 0.03,
\end{eqnarray}
for $P_2$ and $P_{\rm long}$, respectively.

As outlined in Appendix~\ref{appendix}, a phase-offset $\delta$ of the lower of the two frequencies ($\nu_2 = 1/P_2$ in our case) giving rise to a beating results in the long-term modulation to be offset in phase by the negative value $-\delta$. This explains the different sign in the offsets and the fact that the sum of the two observed phase-offsets
\begin{equation}
  \Delta\phi_0(P_2) + \Delta\phi_0(P_{\rm long}) = -0.06 \pm 0.04
\end{equation}
is not significantly different from zero.

\section{Comparison with previous results}

In this section we compare our results of two long-term cycles of \fermilat{} data with results that \citet[][\Xing{} hereafter]{Xing2017} obtained by analysing a \fermilat{} data set covering a very similar time span. In particular we discuss their results of a peak shift and of a dip.

In our Fig.~\ref{fig:FermiLAT_fold} we plot periastron $\Phi = 0.0 - 0.5$ and apastron $\Phi = 0.5 - 1.0$ data for the energy range 0.1 - 3 GeV. \Xing{} show in the top panel of their Fig.~5 a plot of a subset for periastron, $\Phi = 0.1 - 0.4$, and a subset for apastron, $\Phi = 0.6 - 0.9$ for data $\leq$ 5.5 GeV and for  a slightly different long-term period, i.e. 1667\,d, instead of our 1659\,d (compatible within their uncertainty). Despite the slightly different periods used for the folding, energy interval and orbital phase intervals, our plots and the plot by \Xing{} are consistent. They show the absence of long-term modulation at periastron and the presence of a long-term modulation at apastron, peaking in Fig.~5 top right panel in \Xing{} at superorbital phase about $\Theta \sim 0.15$ and in the bottom panel of our Fig.~\ref{fig:FermiLAT_fold} around $\Theta \sim 0.2$.

Besides, \Xing{} show in the right column of their Fig.~3 data of the energy range 0.1-300\,GeV folded on their long-term period of 1667\,d for small subsets of data: a single bin of orbital phase $\Phi = 0.1$. The apastron data (right column of their Fig.~3) indeed show a significant folding and in each fit the peak results constant at the same superorbital phase $\Theta = 0.16 \pm 0.03$ (from their Table~3). This value is well consistent with the peak at $\Theta \sim 0.15$ of their Fig.~5, top right panel, as discussed above, and with our value of $\Theta \sim 0.2$. The constant value of the peak phase for apastron data is shown in the bottom panel of Fig.~4 in \Xing{}. This means that for apastron data there is no shift of the peak phase.

Different is the result for periastron data. At periastron there is no long-term spectral feature in our spectrum shown in Fig.~\ref{fig:FermiLAT_SC}\,a, neither a folding in both our Fig.~\ref{fig:FermiLAT_fold}\,c nor in the top left panel of Fig.~5 in \Xing{}. They fit the data (their Fig.~3, left column) nevertheless with a sine function and discuss the found large $\chi^2$. The fact that the resulting peak phase appears at different phases going from $\Theta = 0.4$ to $\Theta = 1.0$, i.e., their called ``shift'' (their Table~3 and the bottom panel of their Fig.~4) just reflects indeed the noise-dominated fit.

Finally, \Xing{} show in their Fig.~3, left column, that in the periastron data folded on a period of 1667\,d there appears a ``dip'' at superorbital phase $\Theta = 0.65$, the same in their Fig.~5, top right. However, the dip occurs at only one bin. Its noisy nature is clear when it disappears in data folded with a slightly different period: our Fig.~5\,c, folded with 1659\,d does not show any dip at about $\Theta = 0.65$. 

\section{Discussion} \label{sect:discussion}

The observed flux density from a relativistic jet \citep{Mirabel1999} is the product of an intrinsically variable jet that depends on the orbital period\ $P_1$ and Doppler boosting (DB) towards the observer that depends on the jet precession period\ $P_2$. In the physical model for \lsi{} \citep{Massi2014, Jaron2016} this implies
\begin{equation}
  S_{\rm observed}(t) = S_{\rm intrinsic}\left(\Phi_{P_1}(t)\right) \times DB\left(\Phi_{P_2}(t)\right).
\end{equation}
The DB factor depends on the jet velocity and as discussed in \citet{Jaron2016} the low velocity of the jet at periastron, Comptoinized before the acceleration region, explains the dependence of GeV emission at periastron on $P_1$ only and the lack of any radio outburst (i.e., electrons ejected in the strong stellar UV radiation suffer catastrophic inverse Compton losses). At apastron  the larger distance from the stellar radiation field, and consequently lower inverse Compton losses for the electrons, results in the radio and GeV emission to be both affected by the $DB(\Phi_{P_2}(t))$ term. Our result of an offset for $P_2$ between radio and GeV emission implies a different angle with respect to the line of sight for the radio (15~GHz) jet and GeV jet. In fact a different angle induces a different DB and therefore a different phase in the long term modulation. The maximum amplitude corresponds to an alignement of $P_1$ and $P_2$, i.e., the emitting electrons are ejected at the minimum angle with respect to the line of sight.

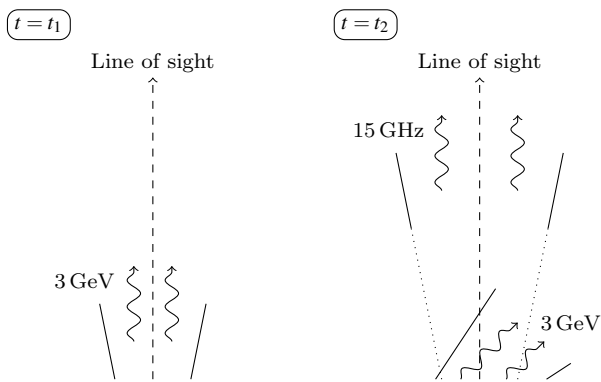
\begin{figure}
  \begin{tikzpicture}
    \fill (-1.5,4.7) node[draw, rounded corners] {$t = t_1$};
    \draw[->, dashed] (0,0) -- (0,4);
    \draw (0,4.2) node {Line of sight};
    \draw (-.5,0) -- (-.7,1);
    \draw (.5,0) -- (.7,1);
    \draw[->, decorate, decoration={snake, post length=.5mm}] (0.25,0.5) -- (0.25,1.5);
    \draw[->, decorate, decoration={snake, post length=.5mm}] (-0.25,0.5) -- (-0.25,1.5);
    \draw (-0.9,1.3) node {3\,GeV};
  \end{tikzpicture}
  \hspace{36pt}
  \begin{tikzpicture}
    \fill (-1.5,4.7) node[draw, rounded corners] {$t = t_2$};
    \draw[->, dashed] (0,0) -- (0,4);
    \draw (0,4.2) node {Line of sight};
    \draw[dotted] (-.5,0) -- (-.9,2);
    \draw[dotted] (.5,0) -- (.9,2);
    \draw (-.9,2) -- (-1.1,3);
    \draw (.9,2) -- (1.1,3);
    \draw[->, decorate, decoration={snake, post length=.5mm}] (0.5,2.5) -- (0.5,3.5);
    \draw[->, decorate, decoration={snake, post length=.5mm}] (-0.5,2.5) -- (-0.5,3.5);
    \draw (-1.2,3.3) node {15\,GHz};
    \draw (-.58,0) -- (.21,1.20);
    \draw (.88,0) -- (1.20,0.21);
    \draw[->, decorate, decoration={snake, post length=.5mm}] (-0.25,0) -- (0.5,.75);
    \draw[->, decorate, decoration={snake, post length=.5mm}] (0.35,0) -- (0.85,.5);
    \draw (1.2,.75) node {3\,GeV};
  \end{tikzpicture}
  \caption{
    Sketch of a scenario in which the GeV emission is produced upstream (i.e., earlier in time) in the jet as compared to the 15\,GHz radio emission. Left: At time $t_1$ the GeV emission is emitted into the direction of the line of sight. Right: At time $t_2$ this population of electrons has cooled and emits in radio at 15\,GHz. The new population of electrons, now emitting at GeV energies, are ejected into a different direction because of jet precession. The difference between $t_1$ and $t_2$ translates to a difference in phase when folding the radio and the GeV data on the precession period.
  }
  \label{fig:sketch}
\end{figure}

A model analogous to the scenario in which \citet{Lisakov2017} explain observed time lags between radio and \gammaray{} emission in active galactic nucleus\ 3C\,273, could qualitatively explain the here observed phase-shift between the radio and \gammaray{} emission, as sketched in Fig.~\ref{fig:sketch}. Particles are ejected into the jet in one direction, and at time $t_1$ these high-energy particles emit \gammaray{} emission via the inverse Compton process (left part of the sketch). Here the bulk flow is aligned with the line-of-sight (LoS) and the \gammaray{} emission is maximally Doppler boosted\footnote{In general the maximum Doppler boosting occurs when the bulk motion encloses the smallest angle with the line-of-sight and is not necessarily perfectly aligned.}. Later (right part of the sketch), those particles have moved further down the jet with the bulk velocity. The emission region has expanded and cooled and now those particles emit 15\,GHz synchrotron emission. This bulk flow is still aligned with the LoS and now the 15\,GHz emission is maximally boosted. At the same time new particles that have been ejected into a different direction due to jet precession emit \gammaray{}s at time $t_2$. But now the bulk flow of this new population is less aligned with the LoS and the emission is less boosted. Imagining this as a continuous process (and not discrete as in the simplified sketch), naturally leads to the phase-offset in the Doppler boosting ($P_2$) for radio and \gammaray{}s, and consequently also for $P_{\rm long}$. The implications are:
\begin{enumerate}
\item{
  A displacement of the \gammaray{} and radio emission regions along the jet naturally explains the phase-offset as a phase-offset in the Doppler boosting.
}
\item{
  The phase-shift relates to a distance between the \gammaray{} emission region and the 15\,GHz emission region along the jet.
}
\end{enumerate}

\section{Conclusions}

The X-ray binary \lsi{} features highly periodic emission all over the electromagntic spectrum, one period being a long-term modulation of $P_{\rm long} = 1659$\,d \citep[e.g.,][]{Massi2016}. The \fermilat{} monitoring the entire sky in the GeV since 2008 \citep{Atwood2009}, has now completed two of these long-term cycles of \lsi{}. We performed timing analysis on the \fermilat{} light curve and the simultanous radio data at 15~GHz obtained by long-term OVRO monitoring. We were aimed to compare the  characteristics of the radio and GeV emission. These are our conclusions.
\begin{enumerate}
\item{
  GeV emission covers the whole orbit whereas radio emission is confined only around apastron (Fig.~6), confirming the result of \citet{Jaron2016}.
}
\item{
  Whereas GeV emission around periastron is only modulated with $P_1$, both orbital period $P_1$ and precession period $P_2$ are present in both radio and apastron GeV data. This also confirms the result of \citet{Jaron2016}. 
}
\item{
  The here presented analysis yields the new result that when folding data with $P_2$ there is a phase-offset between radio and GeV data of $\Delta\Phi(P_2) \approx 0.2$.
}
\item{
  The long-term modulation ($P_{\rm long}$) of the flux is present in both radio and apastron GeV data. When data are folded with this period there is an offset between radio GeV data of $\Delta(\Theta)\sim 0.26$.
}
\item{
  The different sign of the offsets of $P_2$ and $P_{\rm long}$ is as predicted from the beating (see Appendix~\ref{appendix}).
}
\item{
  A model as outlined in Sect.~\ref{sect:discussion} is able to qualitatively explain the here observed phase-shift between the radio and GeV emission from \lsi{}, implying that the GeV emission region is upstream from the site of radio emission. Further long-term monitoring of \lsi{} at multiple wavelengths and physical modelling are necessary to quantitatively test this hypothesis.
}
\end{enumerate}

\section*{Acknowledgements}

We thank Eduardo Ros for reading the manuscript. FJ thanks Walter Alef and Helge Rottmann for granting computing time at the MPIfR correlator cluster.
This work has made use of public \textit{Fermi} data obtained from the High Energy Astrophysics Science Archive Research Center (HEASARC), provided by NASA Goddard Space Flight Center.
The OVRO 40~m Telescope Monitoring Program is supported by NASA under awards NNX08AW31G and NNX11A043G, and by the NSF under awards AST-0808050 and AST-1109911.







\appendix

\section{Beating and phase-offset} \label{appendix}

When a signal $f(t)$ is modulated by two frequencies, $\nu_1$ and $\nu_2$, which are only slightly different from each other, the resulting interference pattern is called a \emph{beating}. In the following we assume $\nu_1$ to be the slightly larger frequency, i.e., $\nu_1 \gtrsim \nu_2$. Using the convention $\omega = 2\pi\nu$ we examine the sum of two sine functions,
\begin{eqnarray}
  f_1(t) & = & \sin\omega_1t + \sin\omega_2t\nonumber\\
  & = & 2\sin\left(\frac{\omega_1 + \omega_2}{2}t\right)\cos\left(\frac{\omega_1 - \omega_2}{2}t\right),
\end{eqnarray}
which can be rewritten as a sine function oscillating at the average frequency $\nu_{\rm average} = (\nu_1 + \nu_2)/2$, slowly modulated by a cosine term with a frequency of $\nu_{\rm cos} = (\nu_1 - \nu_2)/2$. The beat frequency is defined as the frequency of the envelope of the interference pattern, $\nu_{\rm beat} = 2\nu_{\rm cos} = \nu_1 - \nu_2$, and is oscillating at twice the frequency of the cosine term. This is what in \lsi{} corresponds to $\nu_{\rm long} = P_{\rm long}^{-1}$.

A phase-shift $\delta$ of the sine wave oscillating at $\nu_2$ results in
\begin{eqnarray}
  && f_2(t) = \sin\omega_1t + \sin\left(\omega_2t + \delta\right)\nonumber\\
  && = 2\sin\left(\frac{\omega_1 + \omega_2}{2}t + \frac{\delta}{2}\right)\cos\left(\frac{\omega_1 - \omega_2}{2}t - \frac{\delta}{2}\right),
\end{eqnarray}
which shows that the phase-shift $\delta$ affecting the larger frequency $\nu_2$ has the effect of phase-shifting the slowly oscillating cosine term by $-\delta/2$, i.e., in the opposite direction. The envelope, however, which has a frequency of $\nu_{\rm beat} = 2\nu_{\cos}$ is shifted by $-\delta$, which means it experiences the same phase-shift as the sine wave at $\nu_2$ but in the opposite direction.

\bsp	
\label{lastpage}
\end{document}